\documentclass[aps,prd,preprint,amsmath,amssymb,
nofootinbib,eqsecnum,showpacs,preprint,tightenlines]{revtex4}
\usepackage{epsfig}
\usepackage{rotating}
\usepackage{bm}
\newcommand{\be}{\begin{equation}}
\newcommand{\ee}{\end{equation}}
\newcommand{\bea}{\begin{eqnarray}}
\newcommand{\eea}{\end{eqnarray}}
\begin{document}
\title{Casimir Energies and Pressures for $\delta$-function Potentials}
\author{Kimball A. Milton}
\email{milton@nhn.ou.edu}
\homepage{www.nhn.ou.edu/
\affiliation{Department of Physics and Astronomy, University of Oklahoma,
Norman, OK 73019-2061}

\date{\today}

\begin{abstract}
The Casimir energies and pressures for a massless scalar field
associated with $\delta$-function potentials in $1+1$ and $3+1$
dimensions are calculated.  For parallel
plane surfaces, the results are finite, coincide with the pressures
associated with Dirichlet planes in the limit of strong coupling, and for weak
coupling do not possess a power-series expansion in $1+1$ dimension.  
The relation between Casimir energies and Casimir pressures is clarified,
and the former are shown to involve surface terms.
The Casimir energy for a $\delta$-function spherical shell in $3+1$ dimensions
has an expression that reduces to the familiar result for a Dirichlet shell
in the strong-coupling limit.  However, the Casimir energy for finite coupling
possesses a logarithmic divergence first appearing in third order in
the weak-coupling expansion, which seems unremovable.  The corresponding
energies and pressures for a derivative of a $\delta$-function potential 
for the same spherical geometry generalizes the TM contributions  of
electrodynamics.  Cancellation of divergences can occur between the TE 
($\delta$-function) and TM (derivative of $\delta$-function) Casimir energies. 
These results clarify recent discussions in the literature. 
\end{abstract}

\maketitle
\section{Introduction}
Since the inception of quantum mechanics, 
divergences associated with zero-point energy have
caused much confusion.  One way to deal with them was to simply define
them away.  This view, however, appears to be untenable, in view of the
observable consequence of zero-point fluctuations in the Casimir effect,
well probed experimentally  \cite{Bordag:2001qi,miltonbook}.
Calculations of such forces, and of the
associated energies, are generically plagued with infinities.  One modern
consensus is that Casimir forces between distinct bodies may be unambiguously
computed, while self-stresses (the very concept of which is only somewhat
hazily understood) are typically divergent.  There are some famous 
counterexamples: Boyer's result for the Casimir energies of a perfectly
conducting spherical shell \cite{boyersphere}, 
and its generalizations to other geometries \cite{deraadcyl},
dimensions \cite{benmil,mildim}, and fields \cite{johnson,miltonfermion}.  
Even situations which possess manifestly divergent
energies, such as a dielectric ball
\cite{miltonballs}, possess unambiguous finite dilute
limits \cite{bmm,barton}, attributable to van der Waals forces \cite{sonokm}.

Although these difficulties have been known since at least 1979
\cite{miltonballs,deutsch,candelas,candelas2}, recently
they were rediscovered and reexamined in a series of papers by the MIT group
\cite{graham,Graham:2002fw,graham2,Jaffe:2003ji,Graham:2003ib,Weigel:2003tp}.
Perhaps more heat that light has been generated by some of the recent 
discussions.  It is the aim of the present paper to put the discussion on
a somewhat clearer footing by examining Casimir energies and pressures
of massless
scalar fields in a $\delta$-function potential background.  (This is what
the MIT group now refer to a the ``sharp'' limit \cite{Graham:2003ib}.)  
It is then possible
to solve the problem exactly, and study how the result depends on the
strength of the coupling. Although such calculations have been presented
by the MIT group \cite{graham2,Graham:2002fw,Graham:2003ib,Weigel:2003tp}
based on the summation of Feynman diagrams,  they seem not to have
appreciated that Casimir energies for such potentials
 were first computed by the Leipzig group.  The first calculations
with planar $\delta$-function potentials were those of Bordag et 
al.~\cite{hennig}, who found equivalent expressions for the
Casimir energies given later in Refs.~\cite{Graham:2003ib,Weigel:2003tp}
The corresponding spherical problem was studied first by Bordag et al.
\cite{bkv}, who found a nonvanishing second heat kernel coefficient,
indicating that the Casimir energy was divergent in third order in the
coupling.  After a perhaps dubious renormalization, Scandurra 
\cite{Scandurra:1998xa}
extracted the finite part.  Recently, Barton \cite{barton03} has carried
out related calculations, modeling a Fullerine molecule to control
and physically interpret the divergences, and examining the TE and TM
electromagnetic modes, with conclusions not too dissimilar from those
of the MIT group.

Although, therefore, this model seems quite well-studied, it is perhaps
worthwhile to re-examine it in what I consider the most physically transparent
Green's function approach, to see if some clarity can be brought to what
seems at present a rather confused situation.\footnote{Barton \cite{barton03}
refers to my approach as ``older methods,'' but he employs methods of Debye
going back to early in the previous century, and other classic techniques.
I certainly feel in good company if I use the propagation functions invented
by Green, as well as Debye expansions.}
 In so doing, we shall clarify the discussion
of the perturbative expansion, and learn that it is only the strong-coupling
limit of the spherical Casimir energy that possess a finite self-stress, unless
cancellations can occur between TE and TM modes (which certainly do occur
in the strong coupling limit).

This paper is laid out as follows.  In the next section, we find the Casimir
pressure for a massless scalar interacting with two $\delta$-function potentials
in one spatial dimension. (Equivalently, this is a spherical geometry in one
dimension.)
 The pressure is completely finite, but is nonanalytic
in the coupling for weak coupling.  The Casimir energy receives contributions
from the boundaries (surface terms). The generalization to $\delta$-function
planes in three dimensions is immediate, and given in Sec.~\ref{Sec2.5}.
Sec.~\ref{Sec3} presents the corresponding
calculation for the Casimir energy of a massless scalar interacting with a
spherical $\delta$-function shell.  That resulting expression,
in the strong-coupling limit,  reduces to the
standard one for a Dirichlet shell, yielding a finite self-energy 
\cite{Milton:2002vm}.  However,
for any finite coupling, the expression possesses an irremovable logarithmic
divergence, which first appears in third-order in the weak-coupling
expansion \cite{bkv,Graham:2003ib,Weigel:2003tp}, although in second order,
as noted previously \cite{Milton:2002vm}, the energy is finite.
Section \ref{Sec3.5} presents the Casimir energy and pressure for a 
spherical derivative of a $\delta$-function potential, which, in the strong
coupling limit, corresponds to the TM modes of electrodynamics.  (The 
Dirichlet modes computed in Sec.~\ref{Sec3} correspond to the TE modes.)
Concluding remarks are offered in Sec.~\ref{Sec4}.

\section{$1+1$ dimensions}
\label{Sec1}
We consider a massive scalar field (mass $\mu$)
 interacting with two $\delta$-function
potentials, one at $x=0$ and one at $x=a$, which has an interaction
Lagrange density
\be
\mathcal{L}_{\rm int}=-\frac12\frac{\lambda}a\delta(x)\phi^2(x)
-\frac12\frac{\lambda'}a\delta(x-a)\phi^2(x),
\ee
where we have chosen the coupling constants $\lambda$ and $\lambda'$
to be dimensionless.  (But see the following.)
  The Casimir energy for this
situation may be computed in terms of the Green's function $G$,
\be
G(x,x')=i\langle T\phi(x)\phi(x')\rangle,
\ee
which has a time Fourier transform,
\be
G(x,x')=\int\frac{d\omega}{2\pi}e^{-i\omega(t-t')}g(x,x';\omega),
\ee
which in turn satisfies
\be
\left[-\frac{\partial^2}{\partial x^2}+\kappa^2+\frac{\lambda}a\delta(x)
+\frac{\lambda'}a\delta(x-a)\right]g(x,x')=\delta(x-x').
\ee
Here $\kappa^2=\mu^2-\omega^2$.
This equation is easily solved, with the result
\begin{subequations}
\label{gee}
\bea
g(x,x')&=&\frac1{2\kappa}e^{-\kappa|x-x'|}+
\frac1{2\kappa\Delta}\Bigg[\
\frac{\lambda\lambda'}{(2\kappa a)^2}2\cosh\kappa|x-x'|\nonumber\\
&&\quad\mbox{}-\frac{\lambda}{2\kappa a}\left(1+\frac{\lambda'}
{2\kappa a}\right)e^{2\kappa a}
e^{-\kappa(x+x')}-\frac{\lambda'}{2\kappa a}\left(1+\frac{\lambda}
{2\kappa a}\right)e^{\kappa(x+x')}
\label{gin}
\eea
for both fields inside, $0<x,x'<a$, while if both field points are outside,
$a<x,x'$, 
\be
g(x,x')=\frac1{2\kappa}e^{-\kappa|x-x'|}+\frac1{2\kappa\Delta}e^{-\kappa
(x+x'-2a)}\left[-\frac{\lambda}{2\kappa a}\left(1-\frac{\lambda'}
{2\kappa a}\right)
-\frac{\lambda'}{2\kappa a}\left(1+\frac{\lambda}
{2\kappa a}\right)e^{2\kappa a}\right].
\label{gout}
\ee
For $x,x'<0$,
\be
g(x,x')=\frac1{2\kappa}e^{-\kappa|x-x'|}+\frac1{2\kappa\Delta}e^{-\kappa
(x+x'-2a)}\left[-\frac{\lambda}{2\kappa a}\left(1+\frac{\lambda'}
{2\kappa a}\right)
-\frac{\lambda'}{2\kappa a}\left(1-\frac{\lambda}
{2\kappa a}\right)e^{2\kappa a}\right].
\label{gleft}
\ee
\end{subequations}
Here, the denominator is
\be
\Delta=\left(1+\frac{\lambda}{2\kappa a}\right)\left(1+\frac{\lambda'}
{2\kappa a}\right)e^{2\kappa a}-\frac{\lambda\lambda'}{(2\kappa a)^2}.
\ee
Note that in the strong coupling limit we recover the familiar results, 
for example, inside
\be
\lambda,\lambda'\to\infty:
\quad g(x,x')\to-\frac{\sinh\kappa x_<\sinh\kappa(x_>-a)}
{\kappa\sinh\kappa a}.
\ee

We can now calculate the force on one of the $\delta$-function points by
calculating the discontinuity of the stress tensor,  obtained from the
Green's function by
\be
\langle T^{\mu\nu}\rangle=\left(\partial^\mu\partial^{\nu\prime}-\frac12
g^{\mu\nu}\partial^\lambda\partial'_\lambda\right)
\frac1iG(x,x')\bigg|_{x=x'}.
\ee
Writing 
\be
\langle T^{\mu\nu}\rangle=\int\frac{d\omega}{2\pi} t^{\mu\nu},
\ee
we find inside
\bea
t_{xx}&=&\frac1{2i}(\omega^2+\partial_x\partial_{x'})g(x,x')
\bigg|_{x=x'}\nonumber\\
&=&\frac1{4i\kappa\Delta}\Bigg\{(2\omega^2-\mu^2)\left[\left(1+\frac{\lambda}
{2\kappa a}\right)\left(1+\frac{\lambda'}{2\kappa a}\right)
e^{2\kappa a}+\frac{\lambda\lambda'}{(2\kappa
a)^2}\right]\nonumber\\
&&\quad\mbox{}-\mu^2\left[\frac{\lambda}{2\kappa a}\left(1+\frac{\lambda'}
{2\kappa a}\right)
e^{-2\kappa(x- a)}+\frac{\lambda'}{2\kappa a}\left(1+\frac{\lambda}
{2\kappa a}\right)e^{2\kappa x}\right]\Bigg\}.
\label{parstin}
\eea
Let us henceforth simplify the considerations by taking the massless limit,
$\mu=0$.  Then the stress tensor just to the left of the point $x=a$ is
\be
t_{xx}\bigg|_{x=a-}=-\frac\kappa{2i}\left\{
1+2\left[\left(\frac{2\kappa a}\lambda+1\right)\left(\frac{2\kappa
a}{\lambda'}+1\right)e^{2\kappa a}-1\right]^{-1}\right\}.
\label{stressout}
\ee
From this we must subtract the stress just to the right of the point at
$x=a$, obtained from Eq.~(\ref{gout}), which turns out to be in the massless
limit
\be
t_{xx}\bigg|_{x=a+}=-\frac\kappa{2i},
\ee
which just cancels the 1 in braces in Eq.~(\ref{stressout}).
Thus the force on the point $x=a$ due to the quantum fluctuations in the scalar
field is given by the simple, finite expression
\be
F=\langle T_{xx}\rangle\bigg|_{x=a-}-\langle T_{xx}\rangle\bigg|_{x=a+}
=-\frac1{4\pi a^2}\int_0^\infty
dy\,y\,\frac1{(y/\lambda+1)(y/\lambda'+1)e^y-1}.
\label{cepoints}
\ee
This reduces to the well-known, L\"uscher result \cite{luscher,luscher2}
in the limit $\lambda,\lambda'\to
\infty$,
\be
\lim_{\lambda=\lambda'\to\infty}F=-\frac\pi{24a^2},
\ee
and for $\lambda=\lambda'$ is plotted in Fig.~\ref{fig1}.
\begin{figure}
\begin{center}
\begin{turn}{270}
\epsfig{figure=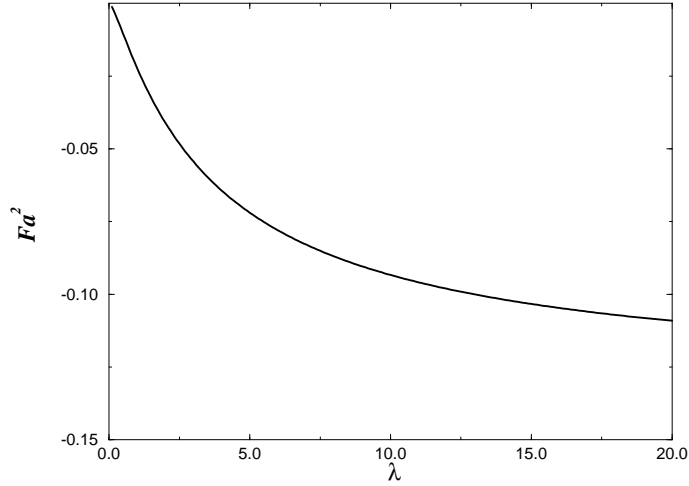,height=10cm}
\end{turn}
\end{center}
\caption{Casimir force between two $\delta$-function points having strength
$\lambda$ and separated by a distance $a$.}
\label{fig1}
\end{figure}

We can also compute the energy density.  In this simple 
massless case, the calculation
appears identical, because $t_{xx}=t_{00}$ (conformal invariance).  
The energy density is constant [Eq.~(\ref{parstin}) with $\mu=0$]
and subtracting from it the $a$-independent part that would be present if
no potential were present, we immediate see that the total energy is
$E=Fa$, so $F=-\partial E/\partial a$.  This result  differs
significantly from that given in 
Refs.~\cite{Graham:2002fw,graham2,Jaffe:2003ji}, which is a 
divergent expression in the massless limit, not  transformable into the 
expression found by this naive procedure.  However, that result may be easily
derived from the following expression for the total energy,
\bea
E&=&\int(d\mathbf{r})\,\langle T^{00}\rangle=\frac1{2i}\int(d\mathbf{r})
(\partial^0\partial^{\prime0}-\nabla^2)G(x,x')\bigg|_{x=x'}\nonumber\\
&=&\frac1{2i}\int(d\mathbf{r})\int\frac{d\omega}{2\pi}2\omega^2G(\mathbf{r,r}).
\label{casenergy}
\eea  Integrating over the Green's functions in the three regions,
given by Eqs.~(\ref{gin}), (\ref{gout}), and (\ref{gleft}), we obtain for
$\lambda=\lambda'$,
\be
E=\frac1{4\pi a}\int_0^\infty dy\frac1{1+y/\lambda}-\frac1{4\pi a}
\int_0^\infty dy\,y\frac{1+2/(y+\lambda)}{(y/\lambda+1)^2e^y-1},
\label{11energy}
\ee
where the first term is regarded as an irrelevant constant ($\lambda/a$ is
constant), and the second is the same as that given by Eq.~(70) of
Ref.~\cite{graham2} upon integration by parts.

The origin of this discrepancy is the existence of a surface contribution
to the energy.  Because $\partial_\mu T^{\mu\nu}=0$, we have, for a region
$V$ bounded by a surface $S$,
\be
0=\frac{d}{dt}\int_V (d\mathbf{r}) T^{00}+\oint_S dS_i T^{0i}.
\ee
Here $T^{0i}=\partial^0\phi\partial^i\phi$, so we conclude that there is
an additional contribution to the energy,
\begin{subequations}
\bea
E_s&=&-\frac1{2i}\int d{\bf S}\cdot\mbox{\boldmath{$\nabla$}}
 G(x,x')\bigg|_{x'=x}
\label{es1}\\
&=&-\frac1{2i}\int_{-\infty}^\infty\frac{d\omega}{2\pi}\sum\frac{d}{dx}
g(x,x')\bigg|_{x'=x},
\label{es2}
\eea
\end{subequations}
where the derivative is taken at the boundaries (here $x=0$, $a$) in the
sense of the outward normal from the region in question.  When this surface
term is taken into account the extra terms in Eq.~(\ref{11energy}) 
are supplied.  The integrated formula (\ref{casenergy}) 
automatically builds in this 
surface contribution, as the implicit surface term in the integration
by parts.  (These terms are slightly unfamiliar because they do not arise
in cases of Neumann or Dirichlet boundary conditions.)  See Fulling 
\cite{Fulling:2003zx} for further discussion.

It is interesting to
consider the behavior of the force or energy for small coupling $\lambda$.
It is clear that, in fact, Eq.~(\ref{cepoints}) is not analytic at $\lambda=0$.
(This reflects an infrared divergence in the Feynman diagram calculation.)
If we expand out the leading $\lambda^2$ term we are left with a divergent
integral.  A correct asymptotic evaluation leads to the behavior
\be
F\sim \frac{\lambda^2}{4\pi a^2}\left(\ln 2\lambda+\gamma\right),\quad 
E\sim-\frac{\lambda^2}{4\pi a}(\ln2\lambda+\gamma-1),\quad
\lambda\to 0.\label{paragsmall}
\ee
This behavior indeed was anticipated in earlier perturbative analyses.
In Ref.~\cite{Milton:2002vm} the general result was given for the Casimir
energy for a $D$ dimensional spherical $\delta$-function potential
(a factor of $1/4\pi$ was inadvertently omitted)
\be
E=-2^{-1-2D}\frac{\lambda^2}{\pi a}
\frac{\Gamma\left(\frac{D-1}2\right)\Gamma(D-3/2)
\Gamma(1-D/2)}{[\Gamma(D/2)]^2}.\label{Ed}
\ee
This possesses an infrared divergence as $D\to1$:
\be
E^{(D=1)}=\frac{\lambda^2}{4\pi a}\Gamma(0),
\ee
which is consistent with the nonanalytic behavior seen in Eq.~(\ref{paragsmall}).

\section{Parallel Planes in $3+1$ Dimensions}
\label{Sec2.5}
It is trivial to extract the expression for the Casimir pressure between two
$\delta$ function planes in three spatial dimensions, where the background
lies at $x=0$ and $x=a$.  We merely have to insert into the above a
transverse momentum transform,
\be
G(x,x')=\int\frac{d\omega}{2\pi}e^{-i\omega(t-t')}\int\frac{(d\mathbf{k})}
{(2\pi)^2}e^{i\mathbf{k\cdot(r-r')_\perp}}g(x,x';\kappa),
\ee
where now $\kappa^2=\mu^2+k^2-\omega^2$.  Then $g$ has exactly the same form
as in Eqs.~(\ref{gee}).
The reduced stress tensor is given by, for the massless case,
\be
t_{xx}=\frac12(\partial_x\partial_{x'}-\kappa^2)\frac1i g(x,x')\bigg|_{x=x'},
\ee
so we immediately see that the attractive pressure on the planes is given by
($\lambda=\lambda'$)
\be
P=-\frac1{32\pi^2 a^4}\int_0^\infty dy\,y^3\,\frac1{(y/\lambda+1)^2e^y-1},
\label{31pressure}
\ee
which coincides with the result given in 
Refs.~\cite{Graham:2003ib,Weigel:2003tp}.  

The Casimir energy per unit area again might be expected to be
\be
\mathcal{E}=-\frac1{96\pi^2a^3}\int_0^\infty dy\frac{y^3}{(y/\lambda+1)^2
e^y-1}=\frac13\frac{P}a,
\label{naivee}
\ee
because then $P=-\frac\partial{\partial a}\mathcal{E}$.  In fact, however,
it is straightforward to compute the energy density $\langle T^{00}\rangle$ 
is the three
regions, $z<0$, $0<z<a$, and $a<z$, and then integrate it over $z$ to
obtain the energy/area, which differs from Eq.~(\ref{naivee}) because, now,
there exists transverse momentum. We also must include the surface term
(\ref{es1}), which is of opposite sign, and of double magnitude, to the
$k^2$ term. The net extra term is
\be
\mathcal{E}'=\frac1{48\pi^2a^3}\int_0^\infty dy\,y^2\frac1{1+y/\lambda}
\left[1-\frac{y/\lambda}{(y/\lambda+1)^2e^y-1}\right].
\ee
If we regard $\lambda/a$ as constant (so that the strength of the coupling
is independent of the separation between the planes) we may drop the first,
divergent term here as irrelevant, being independent of $a$,
because $y=2\kappa a$,  and then the
total energy is
\be
\mathcal{E}=-\frac1{96\pi^2a^3}\int_0^\infty dy\,y^3\frac{1+2/(\lambda+y)}
{(y/\lambda+1)^2e^y-1},
\label{31energy}
\ee
which coincides with the massless limit of the energy first found by Bordag
et al.~\cite{hennig}, and given in Refs.~\cite{Graham:2003ib,Weigel:2003tp}.  
As noted in Sec.~\ref{Sec1}, this result may also readily be derived through
use of (\ref{casenergy}).  When differentiated with respect to $a$,
Eq.~(\ref{31energy}), with $\lambda/a$ fixed, yields the pressure 
(\ref{31pressure}).

\section{Three-dimensional Spherical Potential}
\label{Sec3}

We now carry out the same calculation in three spatial dimensions,
with a radially symmetric background
\be
\mathcal{L}_{\rm int}=-\frac12\frac{\lambda}a\delta(r-a)\phi^2(x),
\ee
which would correspond to a Dirichlet shell in the limit $\lambda\to\infty$.
The time-Fourier transformed Green's function satisfies the equation
($\kappa^2=-\omega^2$)
\be
\left[-\nabla^2+\kappa^2+\frac{\lambda}a\delta(r-a)\right]G(\mathbf{r,r'})=
\delta(\mathbf{r-r'}).
\ee
We write $G$ in terms of a reduced Green's function
\be
G(\mathbf{r,r'})=\sum_{lm}g_l(r,r')Y_{lm}(\Omega)Y^*_{lm}(\Omega'),
\ee
where $g_l$ satisfies
\be
\left[-\frac1{r^2}\frac{d}{dr}r^2\frac{d}{dr}+\frac{l(l+1)}{r^2}+\kappa^2
+\frac{\lambda}a\delta(r-a)\right]g_l(r,r')=\frac1{r^2}\delta(r-r').
\label{redgf}
\ee
We solve this in terms of modified Bessel functions, $I_\nu(x)$, $K_\nu(x)$,
where $\nu=l+1/2$, which satisfy the Wronskian condition
\be
I'_\nu(x)K_\nu(x)-K'_\nu(x)I_\nu(x)=\frac1x.
\ee
We solve Eq.~(\ref{redgf}) by requiring continuity of $g_l$ at each
singularity, $r'$ and $a$, and the appropriate discontinuity of the
derivative.  Inside the sphere we then find ($0<r,r'<a$)
\be
g_l(r,r')=\frac1{\kappa r r'}\left[e_l(\kappa r_>)s_l(\kappa r_<)
-\frac{\lambda}{\kappa a}s_l(\kappa r)s_l(\kappa r')\frac{e_l^2(\kappa a)}{1
+\frac{\lambda}{\kappa a}s_l(\kappa a)e_l(\kappa a)}\right].
\label{insphgf}
\ee
Here we have introduced the modified Riccati-Bessel functions,
\be
s_l(x)=\sqrt{\frac{\pi x}2}I_{l+1/2}(x),\quad
e_l(x)=\sqrt{\frac{2 x}\pi}K_{l+1/2}(x).
\ee
Note that Eq.~(\ref{insphgf}) 
reduces to the expected result, vanishing as $r\to a$,
in the limit of strong coupling:
\be
\lim_{\lambda\to\infty} g_l(r,r')=\frac1{\kappa r r'}\left[e_l(\kappa r_>)
s_l(\kappa r_<)-\frac{e_l(\kappa a)}{s_l(\kappa a)}s_l(\kappa r)s_l
(\kappa r')\right].
\ee
When both points are outside the sphere, $r,r'>a$, we obtain a similar
result:
\be
g_l(r,r')=\frac1{\kappa r r'}\left[e_l(\kappa r_>)s_l(\kappa r_<)
-\frac{\lambda}{\kappa a}e_l(\kappa r)e_l(\kappa r')\frac{s_l^2(\kappa a)}{1
+\frac{\lambda}{\kappa a}s_l(\kappa a)e_l(\kappa a)}\right].
\label{outsphgf}
\ee
which similarly reduces to the expected result as $\lambda\to\infty$.

Now we want to get the radial-radial component of the stress tensor
to get the pressure on the sphere, which is obtained by applying the
operator
\be
\partial_r\partial_{r'}-\frac12(-\partial^0\partial^{\prime0}+\bm{\nabla}
\cdot\bm{\nabla'})\to\frac12\partial_r\partial_{r'}-\kappa^2-\frac{l(l+1)}{r^2}
\label{radop}
\ee to the Green's function, where in the last term we have 
averaged over the surface of the sphere.
In this way we find, from the discontinuity of
$\langle T_{rr}\rangle$ across the $r=a$ surface, the net stress
\be
F=\frac{\lambda}{2\pi a^2}\sum_{l=0}^\infty  (2l+1)\int_0^\infty dx\,
\frac{\left(e_l(x)s_l(x)\right)'-\frac{2e_l(x)s_l(x)}x}{1+\frac{\lambda
e_l(x)s_l(x)}x}.
\label{teforce}
\ee

The same result can be deduced by computing the total energy (\ref{casenergy}).
The free Green's function, the first term in Eqs.~(\ref{insphgf}) or 
(\ref{outsphgf}), evidently makes no significant contribution to the energy,
for it gives a term independent of the radius of the sphere, $a$, so we
omit it.  The remaining radial integrals are simply
\begin{subequations}
\label{radints}
\bea
\int_0^x dy\,s_l^2(y)&=&\frac1{2x}\left[\left(x^2+l(l+1)\right)s_l^2
+xs_ls_l'-x^2s_l^{\prime2}\right],\\
\int_x^\infty dy\,e_l^2(y)&=&
-\frac1{2x}\left[\left(x^2+l(l+1)\right)e_l^2
+xe_le_l'-x^2e_l^{\prime2}\right],
\eea
\end{subequations}
where all the Bessel functions on the right-hand-sides of these equations
are evaluated at $x$.  Then using the Wronskian, we find that
the Casimir energy is
\be
E=-\frac{1}{2\pi a}\sum_{l=0}^\infty  (2l+1)\int_0^\infty dx\,x\,
\frac{d}{dx}\ln\left[1+\lambda I_\nu(x)K_\nu(x)\right].
\label{teenergy}
\ee
If we differentiate with respect to $a$, with $\lambda/a$ fixed, we
immediately recover the force (\ref{teforce}).  This expression, upon
integration by parts, coincides with that given by Barton \cite{barton03},
and was first analyzed in detail by Scandurra \cite{Scandurra:1998xa}.
It reduces to the well-known expression for the Casimir energy
of a massless scalar field  inside and
outside a sphere upon which Dirichlet boundary conditions are imposed,
that is, that the field must vanish at $r=a$:
\be
\lim_{\lambda\to\infty}E=-\frac{1}{2\pi a}\sum_{l=0}^\infty (2l+1)\int_0^\infty
dx\,x\,\frac{d}{dx}\ln\left[I_\nu(x)K_\nu(x)\right],\label{dsph}
\ee
because multiplying the argument of the logarithm by a power of $x$ is
without effect, corresponding to a contact term.  Details of the evaluation
of Eq.~(\ref{dsph}) are given in Ref.~\cite{Milton:2002vm}.

The opposite limit is of interest here.  The expansion of the logarithm
is immediate for small $\lambda$.  The first term, of order $\lambda$, 
is evidently
divergent, but irrelevant, since that may be removed by renormalization
of the tadpole graph.  In contradistinction to the claim of 
Refs.~\cite{Graham:2002fw,graham2,Graham:2003ib,Weigel:2003tp}, 
the order $\lambda^2$ term is finite,
as claimed in Ref.~\cite{Milton:2002vm}.  That term is
\be
E^{(\lambda^2)}=\frac{\lambda^2}{4\pi a}
\sum_{l=0}^\infty(2l+1)\int_0^\infty dx\,x
\frac{d}{dx}[I_{l+1/2}(x)K_{l+1/2}(x)]^2.\label{og}
\ee
The sum on $l$ can be carried out using a trick due to Klich \cite{klich}:
The sum rule
\be
\sum_{l=0}^\infty (2l+1)e_l(x)s_l(y)P_l(\cos\theta)=\frac{xy}\rho e^{-\rho},
\ee
where $\rho=\sqrt{x^2+y^2-2xy\cos\theta}$, is squared, and then integrated
over $\theta$, according to
\be
\int_{-1}^1 d\cos\theta P_l(\cos\theta)P_{l'}(\cos\theta)=\delta_{ll'}\frac2
{2l+1}.
\ee
In this way we learn that
\be
\sum_{l=0}^\infty (2l+1)e_l^2(x)s_l^2(x)=\frac{x^2}2\int_0^{4x}\frac{dw}w
e^{-w}.
\ee
Although this integral is divergent, because we did not integrate by parts
in Eq.~(\ref{og}), that divergence does not contribute:
\be
E^{(\lambda^2)}=\frac{\lambda^2}{4\pi a}\int_0^\infty dx\,
\frac12 x \,\frac{d}{dx}\int_0^{4x}
\frac{dw}w e^{-w}=\frac{\lambda^2}{32\pi a},
\label{4.25}
\ee
which is exactly the result (4.25) of Ref.~\cite{Milton:2002vm}, 
which also follows from Eq.~(\ref{Ed}) here.

However, before we wax too euphoric, we recognize that the order $\lambda^3$ 
term
appears logarithmically divergent, just as Refs.~\cite{Graham:2003ib} and 
\cite{Weigel:2003tp} 
claim.  This does not signal a breakdown in perturbation
theory, as the divergence in the $D=1$ calculation did.  Suppose we
subtract off the two leading terms,
\be
E=-\frac1{2\pi a}\sum_{l=0}^\infty(2l+1)\int_0^\infty dx\,x\,\frac{d}{dx}
\left[\ln\left(1+\lambda I_\nu K_\nu\right)-\lambda I_\nu K_\nu+
\frac{\lambda^2}2(I_\nu K_\nu)^2
\right]+\frac{\lambda^2}{32\pi a}.\label{fulle}
\ee
To study the behavior of the sum for large values of $l$, we can use the
uniform asymptotic expansion (Debye expansion),
\be
\nu\gg1:\quad I_\nu(x)K_\nu(x)\sim\frac{t}{2\nu}\left[1+\frac{A(t)}{\nu^2}
+\frac{B(t)}{\nu^4}+\dots\right].
\label{uae}
\ee
Here $x=\nu z$, and $t=1/\sqrt{1+z^2}$.  The functions $A$ and $B$, etc., are
polynomials in $t$.  We now insert this into Eq.~(\ref{fulle}) and expand
not in $\lambda$ but in $\nu$; the leading term is
\be
E^{(\lambda^3)}\sim
\frac{\lambda^3}{24\pi a}\sum_{l=0}^\infty\frac1\nu\int_0^\infty
\frac{dz}{(1+z^2)^{3/2}}=\frac{\lambda^3}{24\pi a}\zeta(1).
\ee
Although the frequency integral is finite, the angular momentum sum is
divergent.  The appearance here of the divergent $\zeta(1)$ seems to
signal an insuperable barrier to extraction of a finite Casimir energy
for finite $\lambda$.

This divergence has been known for many years, and was first calculated
explicitly in 1998 by Bordag et al.~\cite{bkv}, where the second heat kernel
coefficient gave
\be
E\sim \frac{\lambda^3}{48\pi a}\frac1s,\quad s\to0.
\ee
A possible way of dealing with this divergence was advocated in 
Ref.~\cite{Scandurra:1998xa}.

\section{TM Spherical Potential}
\label{Sec3.5}

Of course, the scalar model considered in the previous section is merely
a toy model, and something analogous to electrodynamics is of far more
physical relevance.  There are good reasons for believing that cancellations
occur in general between TE (Dirichlet) and TM (Robin) modes.  Certainly
they do occur in the classic Boyer energy of a perfectly conducting spherical
shell \cite{boyersphere, balian, mildersch}, and the indications
are that such cancellations occur even with imperfect boundary conditions
\cite{barton03}.  Following the latter reference, let us consider the
potential
\be
\mathcal{L}_{\rm int}=\frac12\lambda a\frac1r\frac\partial{\partial r}
\delta(r-a)\phi^2(x).
\label{tmlag}
\ee
In the limit $\lambda\to\infty$ this corresponds to TM boundary conditions.
The reduced Green's function is thus taken to satisfy
\be
\left[-\frac1{r^2}\frac{d}{dr}r^2\frac{d}{dr}+\frac{l(l+1)}{r^2}
+\kappa^2-\frac{\lambda a}r\frac\partial{\partial r}\delta(r-a)\right]g_l(r,r')
=\frac1{r^2}\delta(r-r').
\ee
At $r=r'$ we have the usual boundary conditions, that $g_l$ be continuous, but
that its derivative be discontinuous,
\be
r^2\frac{d}{dr}g_l\bigg|_{r=r'-}^{r=r'+}=-1,
\ee
while at the surface of the sphere the derivative is continuous,
\begin{subequations}
\be
\frac\partial{\partial r}rg_l\bigg|_{r=a-}^{r=a+}=0,
\ee
while the function is discontinuous,
\be
g_l\bigg|_{r=a-}^{r=a+}=-\lambda \frac\partial{\partial r}rg_l.
\ee
\end{subequations}

It is then easy to find the Green's functions.  When both points are
inside the sphere,
\begin{subequations}
\be
r,r'<a:\quad g_l(r,r')=\frac1{\kappa rr'}\left[s_l(\kappa r_<)e_l(\kappa r_>)
-\frac{\lambda\kappa a[e_l'(\kappa a)]^2s_l(\kappa r)s_l(\kappa r')}
{1+\lambda\kappa a e_l'(\kappa a)s_l'(\kappa a)}\right],
\ee
and when both points are outside the sphere,
\be
r,r'>a:\quad g_l(r,r')=\frac1{\kappa rr'}\left[s_l(\kappa r_<)e_l(\kappa r_>)
-\frac{\lambda\kappa a[s_l'(\kappa a)]^2e_l(\kappa r)e_l(\kappa r')}
{1+\lambda\kappa a e_l'(\kappa a)s_l'(\kappa a)}\right].
\ee
\end{subequations}
It is easy to see that these supply the appropriate Robin boundary conditions
in the $\lambda\to\infty$ limit:
\be
\lim_{\lambda\to0}\frac\partial{\partial r}rg_l\bigg|_{r=a}=0.
\ee

The Casimir energy may be readily obtained from Eq.~(\ref{casenergy}),
and we find, using the integrals (\ref{radints}),
\be
E=-\frac1{2\pi a}\sum_{l=0}^\infty (2l+1)\int_0^\infty dx\,x\frac{d}{dx}
\ln\left[1+\lambda x e_l'(x)s_l'(x)\right].
\label{tmenergy}
\ee
The force may be obtained from this by applying $-\partial/\partial a$,
and regarding $\lambda a$ as constant [see Eq.~(\ref{tmlag})], 
or directly, from the Green's function by applying the operator,
\be
t_{rr} =\frac1{2i}\left[\nabla_r\nabla_{r'}-\kappa^2-\frac
{l(l+1)}{r^2}\right]g_l\bigg|_{r'=r},
\ee
which is the same as that in Eq.~(\ref{radop}), except that
\be
\nabla_r=\frac1r\partial_r r,
\ee
appropriate to TM boundary conditions (see Ref.~\cite{mildim}, for example).
Either way, the total stress on the sphere is
\be
F=-\frac\lambda{2\pi a^2}\sum_{l=0}^\infty(2l+1)\int_0^\infty dx\,x^2\frac{[e_l'(x)
s_l'(x)]'}{1+\lambda x e_l'(x)s_l'(x)}.
\ee
The result for the energy (\ref{tmenergy}) is similar, but not identical, to
that given by Barton \cite{barton03}.

Suppose we now combine the TE and TM Casimir energies, Eqs.~(\ref{teenergy}) 
and (\ref{tmenergy}):
\be
E^{\rm TE}+E^{\rm TM}=-\frac1{2\pi a}\sum_{l=0}^\infty (2l+1)\int_0^\infty dx\,x
\frac d{dx}\ln\left[\left(1+\lambda\frac{e_ls_l}x\right)\left(1+\lambda x
e_l's_l'\right)\right].
\label{combenergy}
\ee
In the limit $\lambda\to\infty$ this reduces to the familiar expression
for the perfectly conducting spherical shell \cite{mildersch}:
\be
\lim_{\lambda\to\infty}E=-\frac1{2\pi a}\sum_{l=1}^\infty(2l+1)\int_0^\infty
dx\,x\left(\frac{e_l'}{e_l}+\frac{e_l''}{e_l'}+\frac{s_l'}{s_l}+\frac{s_l''}
{s_l'}\right).
\ee
Here we have, as appropriate to the electrodynamic situation, omitted the
$l=0$ mode. This expression
 yields a finite Casimir energy.  What about finite $\lambda$?  In general,
it appears that there is no chance that the divergence found in the previous
section in order $\lambda^3$ can be cancelled.  But suppose the coupling
for the TE and TM modes are different.  If $\lambda^{\rm TE}\lambda^{\rm 
TM}=4$, a cancellation appears possible.

Let's illustrate this by retaining only the leading terms in the uniform
 asymptotic expansions: ($x=\nu z$)
\be
\frac{e_l(x)s_l(x)}x\sim\frac{t}{2\nu},\quad x e_l'(x)s_l'(x)\sim 
-\frac\nu{2t},\quad\nu\to\infty.
\ee
Then the logarithm appearing in the integral for the energy (\ref{combenergy}) is
approximately
\be
\ln\sim\ln\left(-\frac{\lambda^{\rm TM}\nu}{2t}\right)+\ln\left(1+\frac
{\lambda^{\rm TE}t}{2\nu}\right)+\ln\left(1-\frac{2t}{\lambda^{\rm TM}\nu}
\right).
\ee
The first term here presumably gives no contribution to the energy,
because it is independent of $\lambda$ upon differentiation, and further
we may interpret $\sum_{l=0}^\infty\nu^2=0$ [see Eq.~(\ref{sumzeta})].  
Now if we make the above identification of
the couplings, 
\be
\hat\lambda=\frac{\lambda^{\rm TE}}2=\frac2{\lambda^{\rm TM}},
\label{hatl}
\ee
all the odd powers of $\nu$ cancel out, and
\be
E\sim -\frac1{2\pi a}\sum_{l=0}^\infty(2l+1)\int_0^\infty dx\,x \frac{d}{dx}
\ln\left(1-\frac{{\hat\lambda}^2 t^2}{\nu^2}\right).
\label{leadingtetm}
\ee
The divergence encountered for the TE mode is thus removed, and the power
series is simply twice the sum of the even terms there.  This will be
finite.  Presumably, the same is true if the subleading terms in the uniform
asymptotic expansion are retained.

It is interesting to approximately evaluate Eq.~(\ref{leadingtetm}).
The integral over $z$ may be easily evaluated as a contour integral,
leaving
\be
E\sim-\frac1{a}\sum_{l=0}^\infty \nu^2\left(1-\sqrt{1-\frac{{\hat\lambda}^2}{\nu^2}}
\right).
\ee
This $l$ sum is logarithmic divergent, an artifact of the asymptotic
expansion, since we know the $\lambda^2$ term is finite.  If we expand
the square root for small ${\hat\lambda}^2/\nu^2$, we see that the
$\mathcal{O}({\hat\lambda}^2)$ term vanishes if we interpret the sum as
\be
\sum_{l=0}^\infty \nu^{-s}=(2^s-1)\zeta(s),
\label{sumzeta}
\ee
in terms of the Riemann zeta function.  The leading term is 
$\mathcal{O}({\hat\lambda}^4)$:
\be
E\sim-\frac{{\hat\lambda}^4}{8a}\sum_{l=0}^\infty \frac1{\nu^2}=
\frac{{\hat\lambda}^4\pi^2}
{16a}.
\ee
To recover the correct leading $\lambda$ behavior in (\ref{4.25}) requires
the inclusion of the subleading $\nu^{-2n}$ terms displayed in Eq.~(\ref{uae}).

Much faster convergence is achieved if we consider the results with the
$l=0$ term removed, as appropriate for electromagnetic modes.  Let's illustrate
this for the order $\lambda^2$ TE mode (now, for simplicity, write $\lambda
=\lambda^{\rm TE}$)  Then, in place of the energy (\ref{4.25}) we have
\be
\tilde E^{\lambda^2}=\frac{\lambda^2}{32\pi a}+\frac{\lambda^2}{4\pi a}
\int_0^\infty \frac{dx}{x^2}\sinh^2x \,e^{-2x}=\frac{\lambda^2}a\left(
\frac1{32\pi}+\frac{\ln2}{4\pi}\right)=\frac{\lambda^2}a(0.0651061).
\label{exact}
\ee
Now the leading term in the uniform asymptotic expansion is no longer zero:
\bea
E^{(0)}&=&-\frac1{2\pi a}\sum_{l=1}^\infty (2l+1)\int_0^\infty dx \, x
\frac{d}{dx}\left(-\frac{\lambda^2 t^2}{8\nu^2}\right)\nonumber\\
&=&\frac{\lambda^2}{8\pi a}\sum_{l=1}^\infty \nu^0\left(-\frac\pi2\right)
=\frac{\lambda^2}{16a}=\frac{\lambda^2}a(0.0625),
\eea
which is 4\% lower than the exact answer (\ref{exact}).  The next term
in the uniform asymptotic expansion is
\bea
E^{(2)}&=&-\frac{\lambda^2}{4\pi a}[3\zeta(2)-4]\int_0^\infty dz\,t^2\frac
{t^2-6t^4+5t^5}8\nonumber\\
&=&\frac{\lambda^2}a\left(\frac{3\pi^2}{2048}-\frac{3}{256}
\right)=\frac{\lambda^2}a(0.0027368),
\eea
which reduces the estimate to
\be
E^{(0)}+E^{(2)}=\frac{\lambda^2}a(0.0652368), 
\ee
which is now 0.2\% high.  Going out one more term give
\bea
E^{(4)}&=&-\frac{\lambda^2}{8\pi a}\left[15\zeta(4)-16\right]\int_0^\infty
dz\,t^2\frac{t^4}{16}(7-148t^2+554t^4-708t^6+295t^8)\nonumber\\
&=&-\frac{\lambda^2}a\frac{59\pi^4}{524288}+\frac{\lambda^2}a\frac{177}
{16328}=-\frac{\lambda^2}a(0.000158570),
\eea
and the estimate for the energy is now only 0.04\% low:
\be
E^{(0)}+E^{(2)}+E^{(4)}=\frac{\lambda^2}a(0.06507823).
\ee

We could also make similar remarks about the TM contributions.  However,
evidently there are additional subtleties here, so we will defer further
discussion for an further publication.

\section{Conclusions}
\label{Sec4}

In this paper we have repeated some calculations with ``sharp'' but
not necessarily ``strong'' potentials.  That is, we have computed Casimir
energies in the presence of $\lambda\delta(x-a)$ potentials, in the cases
when the delta function lies on two parallel planes (first considered
in Ref.~\cite{hennig}), and when the
support of the $\delta$ function is a sphere (first considered in 
Ref.~\cite{bkv,Scandurra:1998xa}). 
We have also considered spherical potentials of the form $\lambda\delta'(r-a)
/r$.  For either spherical potential, the approach 
given here yields finite result in all orders, except the third,
provided we make the coupling constant identification (\ref{hatl}) in the
TM case.  That is,
the expression for the energy possess a logarithmic divergence entirely
associated with the order $\lambda^3$ Feynman graph.  This was rediscovered
by Graham et al.~\cite{Graham:2003ib,Weigel:2003tp}, but obscured by
the apparent (spurious) divergence they also claimed to find in order
$\lambda^2$.  The bottom line, however, is that these sharp potentials
yield a divergent Casimir self-stress.

The generalizations drawn in Graham et al.\ papers
\cite{Graham:2002fw,graham2,Graham:2003ib,Weigel:2003tp} 
are, however, perhaps too
strong.  The fact that the $\lambda\to\infty$ limit of the expression for
the energy coincides with that for the Dirichlet shell, does not prove
that the latter is divergent.  It does, however, suggest that that idealization
does not yield the full result for the energy of
a configuration defined by a real material boundary.
This, of course, is no surprise.  It has been recognized since at least
1979 \cite{miltonballs,deutsch}
that constructing a shell from real materials will yield apparent
divergences as the ideal limit is approached, 
so for example, a shell of finite thickness
made of dielectric material will correspond to a divergent Casimir energy.

So the finite Boyer energy \cite{boyersphere}
 for an ideal sphere results from omitting divergent
terms, which may or may not have observable consequences.  (It may be, of
course, that for electromagnetic modes, the divergence found here could
cancel, for which we have provided some evidence.) 
However, what is remarkable, and of some significance, is that
this finite term is unique.  For example, Barton has recently exhibited
a Buckyball model of a conducting spherical shell that possesses various
large energy contributions referring to the material properties of the
shell, but which nevertheless possesses a unique, if subdominant, Boyer term
of order $1/a$ \cite{barton03}.     

It may be useful to compare this situation with a slightly better understood
example, the Casimir energy of a dielectric sphere.  That is certainly
divergent; yet if the divergences are isolated in terms that contribute
to the volume and surface energies, in order $(\epsilon-1)^2$ a unique $1/a$
coefficient emerges \cite{bmm,barton,bkv,hb}, which may be interpretable
as the van der Waals energy \cite{sonokm2}.  
That coefficient diverges in order $(\epsilon-1)^3$ \cite{bkv}.
This fact seems to bear a striking resemblance to the finite Casimir
energy found here in order $\lambda^2$, and the divergence in the next order.
There is also the more than analogous relationship between the finiteness
of the Casimir energy for a dielectric-diamagnetic ball with $\epsilon\mu=1$,
and the finiteness found here when $\lambda^{\rm TE}\lambda^{\rm TM}=4$:
In both cases the divergences separately associated with TE and TM modes
cancel.

There are also extremely interesting issues related to surface divergences
in the local Casimir energy density, which have been discussed recently
by Fulling \cite{Fulling:2003zx}. 
His ideas likely will have bearing on understanding the nature
of the divergences encountered in these problems.

Evidently, there is much work to be done in understanding the nature of
quantum vacuum energy.  It would obviously be of great benefit if it
would be possible to access these questions experimentally.

\begin{acknowledgments}
The author would like to thank Michael Bordag,
Ines Cavero-Pelaez, Ricardo Estrada,
Steve Fulling, Klaus Kirsten,  Kuloth Shajesh, and all the participants of
the recent workshop on Quantum Field Theory Under the Influence of
External Conditions (QFEXT03) for
helpful discussions, and Gabriel Barton for sending me his papers
prior to publication.  I am grateful to
the US Department of Energy for partial financial
support of this research.
\end{acknowledgments}


\begin{thebibliography}{34}
\expandafter\ifx\csname natexlab\endcsname\relax\def\natexlab#1{#1}\fi
\expandafter\ifx\csname bibnamefont\endcsname\relax
  \def\bibnamefont#1{#1}\fi
\expandafter\ifx\csname bibfnamefont\endcsname\relax
  \def\bibfnamefont#1{#1}\fi
\expandafter\ifx\csname citenamefont\endcsname\relax
  \def\citenamefont#1{#1}\fi
\expandafter\ifx\csname url\endcsname\relax
  \def\url#1{\texttt{#1}}\fi
\expandafter\ifx\csname urlprefix\endcsname\relax\def\urlprefix{URL }\fi
\providecommand{\bibinfo}[2]{#2}
\providecommand{\eprint}[2][]{\url{#2}}

\bibitem[{\citenamefont{Bordag et~al.}(2001)\citenamefont{Bordag, Mohideen, and
  Mostepanenko}}]{Bordag:2001qi}
\bibinfo{author}{\bibfnamefont{M.}~\bibnamefont{Bordag}},
  \bibinfo{author}{\bibfnamefont{U.}~\bibnamefont{Mohideen}}, \bibnamefont{and}
  \bibinfo{author}{\bibfnamefont{V.~M.} \bibnamefont{Mostepanenko}},
  \bibinfo{journal}{Phys. Rept.} \textbf{\bibinfo{volume}{353}},
  \bibinfo{pages}{1} (\bibinfo{year}{2001}), \eprint{quant-ph/0106045}.

\bibitem[{\citenamefont{Milton}(2001)}]{miltonbook}
\bibinfo{author}{\bibfnamefont{K.~A.} \bibnamefont{Milton}},
  \emph{\bibinfo{title}{The Casimir Effect: Physical Manifestations of
  Zero-Point Energy}} (\bibinfo{publisher}{World Scientific},
  \bibinfo{address}{Singapore}, \bibinfo{year}{2001}).

\bibitem[{\citenamefont{Boyer}(1968)}]{boyersphere}
\bibinfo{author}{\bibfnamefont{T.~H.} \bibnamefont{Boyer}},
  \bibinfo{journal}{Phys. Rev.} \textbf{\bibinfo{volume}{174}},
  \bibinfo{pages}{1764} (\bibinfo{year}{1968}).

\bibitem[{\citenamefont{{L. L. DeRaad, Jr. and K. A.
  Milton}}(1981)}]{deraadcyl}
\bibinfo{author}{\bibnamefont{{L. L. DeRaad, Jr. and K. A. Milton}}},
  \bibinfo{journal}{Ann. Phys. (N.Y.)} \textbf{\bibinfo{volume}{136}},
  \bibinfo{pages}{229} (\bibinfo{year}{1981}).

\bibitem[{\citenamefont{Bender and Milton}(1994)}]{benmil}
\bibinfo{author}{\bibfnamefont{C.~M.} \bibnamefont{Bender}} \bibnamefont{and}
  \bibinfo{author}{\bibfnamefont{K.~A.} \bibnamefont{Milton}},
  \bibinfo{journal}{Phys. Rev. D} \textbf{\bibinfo{volume}{50}},
  \bibinfo{pages}{6547} (\bibinfo{year}{1994}), \eprint{hep-th/9406048}.

\bibitem[{\citenamefont{Milton}(1997)}]{mildim}
\bibinfo{author}{\bibfnamefont{K.~A.} \bibnamefont{Milton}},
  \bibinfo{journal}{Phys. Rev. D} \textbf{\bibinfo{volume}{55}},
  \bibinfo{pages}{4940} (\bibinfo{year}{1997}), \eprint{hep-th/9611078}.

\bibitem[{\citenamefont{Johnson}(1975)}]{johnson}
\bibinfo{author}{\bibfnamefont{K.}~\bibnamefont{Johnson}},
  \bibinfo{journal}{Acta Phys. Pol.} \textbf{\bibinfo{volume}{B6}},
  \bibinfo{pages}{865} (\bibinfo{year}{1975}).

\bibitem[{\citenamefont{Milton}(1980{\natexlab{a}})}]{miltonfermion}
\bibinfo{author}{\bibfnamefont{K.~A.} \bibnamefont{Milton}},
  \bibinfo{journal}{Phys. Rev. D} \textbf{\bibinfo{volume}{22}},
  \bibinfo{pages}{1444} (\bibinfo{year}{1980}{\natexlab{a}}).

\bibitem[{\citenamefont{Milton}(1980{\natexlab{b}})}]{miltonballs}
\bibinfo{author}{\bibfnamefont{K.~A.} \bibnamefont{Milton}},
  \bibinfo{journal}{Ann. Phys. (N.Y.)} \textbf{\bibinfo{volume}{127}},
  \bibinfo{pages}{49} (\bibinfo{year}{1980}{\natexlab{b}}).

\bibitem[{\citenamefont{{I. Brevik, V. N. Marachevsky, and K. A.
  Milton}}(1999)}]{bmm}
\bibinfo{author}{\bibnamefont{{I. Brevik, V. N. Marachevsky, and K. A.
  Milton}}}, \bibinfo{journal}{Phys. Rev. Lett.} \textbf{\bibinfo{volume}{82}},
  \bibinfo{pages}{3948} (\bibinfo{year}{1999}), \eprint{hep-th/9810062}.

\bibitem[{\citenamefont{Barton}(1999)}]{barton}
\bibinfo{author}{\bibfnamefont{G.}~\bibnamefont{Barton}}, \bibinfo{journal}{J.
  Phys. A} \textbf{\bibinfo{volume}{32}}, \bibinfo{pages}{525}
  (\bibinfo{year}{1999}).

\bibitem[{\citenamefont{Milton and Ng}(1997)}]{sonokm}
\bibinfo{author}{\bibfnamefont{K.~A.} \bibnamefont{Milton}} \bibnamefont{and}
  \bibinfo{author}{\bibfnamefont{Y.~J.} \bibnamefont{Ng}},
  \bibinfo{journal}{Phys. Rev. E} \textbf{\bibinfo{volume}{55}},
  \bibinfo{pages}{4207} (\bibinfo{year}{1997}), \eprint{hep-th/9607186}.

\bibitem[{\citenamefont{Deutsch and Candelas}(1979)}]{deutsch}
\bibinfo{author}{\bibfnamefont{D.}~\bibnamefont{Deutsch}} \bibnamefont{and}
  \bibinfo{author}{\bibfnamefont{P.}~\bibnamefont{Candelas}},
  \bibinfo{journal}{Phys. Rev. D} \textbf{\bibinfo{volume}{20}},
  \bibinfo{pages}{3063} (\bibinfo{year}{1979}).

\bibitem[{\citenamefont{Candelas}(1982)}]{candelas}
\bibinfo{author}{\bibfnamefont{P.}~\bibnamefont{Candelas}},
  \bibinfo{journal}{Ann. Phys. (N.Y.)} \textbf{\bibinfo{volume}{143}},
  \bibinfo{pages}{241} (\bibinfo{year}{1982}).

\bibitem[{\citenamefont{Candelas}(1986)}]{candelas2}
\bibinfo{author}{\bibfnamefont{P.}~\bibnamefont{Candelas}},
  \bibinfo{journal}{Ann. Phys. (N.Y.)} \textbf{\bibinfo{volume}{167}},
  \bibinfo{pages}{257} (\bibinfo{year}{1986}).

\bibitem[{\citenamefont{Graham et~al.}(2002{\natexlab{a}})\citenamefont{Graham,
  Jaffe, and Weigel}}]{graham}
\bibinfo{author}{\bibfnamefont{N.}~\bibnamefont{Graham}},
  \bibinfo{author}{\bibfnamefont{R.~L.} \bibnamefont{Jaffe}}, \bibnamefont{and}
  \bibinfo{author}{\bibfnamefont{H.}~\bibnamefont{Weigel}},
  \bibinfo{journal}{Int. J. Mod. Phys.} \textbf{\bibinfo{volume}{A17}},
  \bibinfo{pages}{846} (\bibinfo{year}{2002}{\natexlab{a}}),
  \eprint{hep-th/0201148}.

\bibitem[{\citenamefont{Graham et~al.}(2003{\natexlab{b}})\citenamefont{Graham,
  Jaffe, Khemani, Quandt, Scandurra, and Weigel}}]{Graham:2002fw}
\bibinfo{author}{\bibfnamefont{N.}~\bibnamefont{Graham}},
  \bibinfo{author}{\bibfnamefont{R.}~\bibnamefont{Jaffe}},
  \bibinfo{author}{\bibfnamefont{V.}~\bibnamefont{Khemani}},
  \bibinfo{author}{\bibfnamefont{M.}~\bibnamefont{Quandt}},
  \bibinfo{author}{\bibfnamefont{M.}~\bibnamefont{Scandurra}},
  \bibnamefont{and} \bibinfo{author}{\bibfnamefont{H.}~\bibnamefont{Weigel}},
  \bibinfo{journal}{Phys. Lett.} \textbf{\bibinfo{volume}{B572}},
  \bibinfo{pages}{196} (\bibinfo{year}{2003}{\natexlab{b}}),
  \eprint{hep-th/0207205}.

\bibitem[{\citenamefont{Graham et~al.}(2002{\natexlab{b}})\citenamefont{Graham,
  Jaffe, Khemani, Quandt, Scandurra, and Weigel}}]{graham2}
\bibinfo{author}{\bibfnamefont{N.}~\bibnamefont{Graham}},
  \bibinfo{author}{\bibfnamefont{R.}~\bibnamefont{Jaffe}},
  \bibinfo{author}{\bibfnamefont{V.}~\bibnamefont{Khemani}},
  \bibinfo{author}{\bibfnamefont{M.}~\bibnamefont{Quandt}},
  \bibinfo{author}{\bibfnamefont{M.}~\bibnamefont{Scandurra}},
  \bibnamefont{and} \bibinfo{author}{\bibfnamefont{H.}~\bibnamefont{Weigel}},
  \bibinfo{journal}{Nucl. Phys. B} \textbf{\bibinfo{volume}{645}},
  \bibinfo{pages}{49} (\bibinfo{year}{2002}{\natexlab{b}}),
  \eprint{hep-th/0207120}.

\bibitem[{\citenamefont{Jaffe}(2003)}]{Jaffe:2003ji}
\bibinfo{author}{\bibfnamefont{R.~L.} \bibnamefont{Jaffe}},
  \bibinfo{journal}{AIP Conf. Proc.} \textbf{\bibinfo{volume}{687}},
  \bibinfo{pages}{3} (\bibinfo{year}{2003}), \eprint{hep-th/0307014}.


\bibitem[{\citenamefont{Graham et~al.}(2003{\natexlab{a}})\citenamefont{Graham,
  Jaffe, Khemani, Quandt, Schroeder, and Weigel}}]{Graham:2003ib}
\bibinfo{author}{\bibfnamefont{N.}~\bibnamefont{Graham}},
  \bibinfo{author}{\bibfnamefont{R.}~\bibnamefont{Jaffe}},
  \bibinfo{author}{\bibfnamefont{V.}~\bibnamefont{Khemani}},
  \bibinfo{author}{\bibfnamefont{M.}~\bibnamefont{Quandt}},
  \bibinfo{author}{\bibfnamefont{O.}~\bibnamefont{Schroeder}},
  \bibnamefont{and} \bibinfo{author}{\bibfnamefont{H.}~\bibnamefont{Weigel}}
  (\bibinfo{year}{2003}{\natexlab{a}}), \eprint{hep-th/0309130}.



\bibitem[{\citenamefont{Weigel}(2003)}]{Weigel:2003tp}
\bibinfo{author}{\bibfnamefont{H.}~\bibnamefont{Weigel}}
  (\bibinfo{year}{2003}), \eprint{hep-th/0310301}.


\bibitem[{\citenamefont{Bordag et~al.}(1992)\citenamefont{Bordag, Hennig, and
  Robaschik}}]{hennig}
\bibinfo{author}{\bibfnamefont{M.}~\bibnamefont{Bordag}},
  \bibinfo{author}{\bibfnamefont{D.}~\bibnamefont{Hennig}}, \bibnamefont{and}
  \bibinfo{author}{\bibfnamefont{D.}~\bibnamefont{Robaschik}},
  \bibinfo{journal}{J. Phys. A} \textbf{\bibinfo{volume}{25}},
  \bibinfo{pages}{4483} (\bibinfo{year}{1992}).

\bibitem[{\citenamefont{{M. Bordag, K. Kirsten, and D.
  Vassilevich}}(1999)}]{bkv}
\bibinfo{author}{\bibnamefont{{M. Bordag, K. Kirsten, and D. Vassilevich}}},
  \bibinfo{journal}{Phys. Rev. D} \textbf{\bibinfo{volume}{59}},
  \bibinfo{pages}{085011} (\bibinfo{year}{1999}), \eprint{hep-th/9811015}.

\bibitem[{\citenamefont{Scandurra}(1999)}]{Scandurra:1998xa}
\bibinfo{author}{\bibfnamefont{M.}~\bibnamefont{Scandurra}},
  \bibinfo{journal}{J. Phys.} \textbf{\bibinfo{volume}{A32}},
  \bibinfo{pages}{5679} (\bibinfo{year}{1999}), \eprint{hep-th/9811164}.

\bibitem[{\citenamefont{Barton}(2003)}]{barton03}
\bibinfo{author}{\bibfnamefont{G.}~\bibnamefont{Barton}}, \bibinfo{journal}{J.
  Phys. A}  (\bibinfo{year}{2003}), \bibinfo{note}{in press}.

\bibitem[{\citenamefont{Milton}(2003)}]{Milton:2002vm}
\bibinfo{author}{\bibfnamefont{K.~A.} \bibnamefont{Milton}},
  \bibinfo{journal}{Phys. Rev. D} \textbf{\bibinfo{volume}{68}},
  \bibinfo{pages}{065020} (\bibinfo{year}{2003}), \eprint{hep-th/0210081}.

\bibitem[{\citenamefont{{M. L\"uscher, K. Symanzik, and P.
  Weisz}}(1980)}]{luscher}
\bibinfo{author}{\bibnamefont{{M. L\"uscher, K. Symanzik, and P. Weisz}}},
  \bibinfo{journal}{Nucl. Phys. B} \textbf{\bibinfo{volume}{173}},
  \bibinfo{pages}{365} (\bibinfo{year}{1980}).

\bibitem[{\citenamefont{{M. L\"uscher}}(1981)}]{luscher2}
\bibinfo{author}{\bibnamefont{{M. L\"uscher}}}, \bibinfo{journal}{Nucl. Phys.
  B} \textbf{\bibinfo{volume}{180}}, \bibinfo{pages}{317}
  (\bibinfo{year}{1981}).

\bibitem[{\citenamefont{Fulling}(2003)}]{Fulling:2003zx}
\bibinfo{author}{\bibfnamefont{S.~A.} \bibnamefont{Fulling}},
  \bibinfo{journal}{J. Phys.} \textbf{\bibinfo{volume}{A36}},
  \bibinfo{pages}{6529} (\bibinfo{year}{2003}), \eprint{quant-ph/0302117}.

\bibitem[{\citenamefont{Klich}(2000)}]{klich}
\bibinfo{author}{\bibfnamefont{I.}~\bibnamefont{Klich}},
  \bibinfo{journal}{Phys. Rev. D} \textbf{\bibinfo{volume}{61}},
  \bibinfo{pages}{025004} (\bibinfo{year}{2000}), \eprint{hep-th/9908101}.

\bibitem[{\citenamefont{Milton et~al.}(1978)\citenamefont{Milton, {L. L.
  DeRaad, Jr.}, and Schwinger}}]{mildersch}
\bibinfo{author}{\bibfnamefont{K.~A.} \bibnamefont{Milton}},
  \bibinfo{author}{\bibnamefont{{L. L. DeRaad, Jr.}}}, \bibnamefont{and}
  \bibinfo{author}{\bibfnamefont{J.}~\bibnamefont{Schwinger}},
  \bibinfo{journal}{Ann. Phys. (N.Y.)} \textbf{\bibinfo{volume}{115}},
  \bibinfo{pages}{388} (\bibinfo{year}{1978}).

\bibitem[{\citenamefont{Balian and Duplantier}(1978)}]{balian}
\bibinfo{author}{\bibfnamefont{R.}~\bibnamefont{Balian}} \bibnamefont{and}
  \bibinfo{author}{\bibfnamefont{B.}~\bibnamefont{Duplantier}},
  \bibinfo{journal}{Ann. Phys. (N.Y.)} \textbf{\bibinfo{volume}{112}},
  \bibinfo{pages}{165} (\bibinfo{year}{1978}).

\bibitem[{\citenamefont{{J. S. H\o ye and I. Brevik}}(2000)}]{hb}
\bibinfo{author}{\bibnamefont{{J. S. H\o ye and I. Brevik}}},
  \bibinfo{journal}{J. Stat. Phys.} \textbf{\bibinfo{volume}{100}},
  \bibinfo{pages}{223} (\bibinfo{year}{2000}), \eprint{quant-ph/9903086}.

\bibitem[{\citenamefont{Milton and Ng}(1998)}]{sonokm2}
\bibinfo{author}{\bibfnamefont{K.~A.} \bibnamefont{Milton}} \bibnamefont{and}
  \bibinfo{author}{\bibfnamefont{Y.~J.} \bibnamefont{Ng}},
  \bibinfo{journal}{Phys. Rev. E} \textbf{\bibinfo{volume}{57}},
  \bibinfo{pages}{5504} (\bibinfo{year}{1998}).

\end{thebibliography}

\end{document}